\documentclass[11pt]{article}%
\usepackage{amsfonts}
\usepackage{amsmath}
\usepackage{amssymb}
\usepackage{graphicx}%
\begin{document}

\title{How classical is the quantum universe?}
\author{Maurice de Gosson\thanks{This work has been financed by the Austrian Research
Agency FWF (Projektnummer P20442-N13). }\\Current address: \textit{Universit\"{a}t Wien}\\\textit{Fakult\"{a}t f\"{u}r Mathematik, NuHAG }\\\textit{Nordbergstrasse 15, AT-1090 Wien}\\maurice.degosson@gmail.com}
\maketitle

\begin{abstract}
We discuss two topics that are usually considered to be exclusively "quantum":
the Schr\"{o}dinger equation, and the uncertainty principle. \ We show (or
rather recall) that the Schr\"{o}dinger equation can be derived from
Hamilton's equations using the metaplectic representation. We also show that
the uncertainty principle, stated in the form of the
Robertson-Schr\"{o}dinger-Heisenberg inequalities can be formulated in
perfectly classical terms using the topological notion of symplectic capacity.

\end{abstract}

\qquad\qquad\qquad\qquad\qquad\qquad\textit{To my parents}

\section{Introduction}

In his recent contribution \cite{Hartle} to the conference \textit{Everett at
50} Hartle observes that

\begin{quotation}
\textquotedblleft...The most striking observable feature of our
indeterministic quantum universe is the wide range of time, place, and scale
on which the deterministic laws of classical physics hold to an excellent
approximation.\textquotedblright\ 
\end{quotation}

In this short essay I will try to complement Hartle's discussion by exposing
two \emph{mathematical} facts that are usually ignored by physicists an, and
which seem to suggest that the quantum universe is, in a sense, far more
classical than it seems at first sight. The first of these facts is about the
Schr\"{o}dinger equation; it should be known by all physicists but experience
shows that this is not the case\footnote{A few years ago, while being a
visiting professor at Yale, I was invited by the mathematical physics group of
a famous University located in new Jersey to give a talk on the topic. My
claim almost triggered a riot among my colleagues physicists!}. The second is
about the uncertainty principle; recent advances in symplectic topology of a
very subtle nature, suggest that there exists an uncertainty principle in
classical mechanics which is formally absolutely similar to a refined version
of the Heisenberg inequalities.

More precisely, I want to point out that:

\begin{itemize}
\item The Schr\"{o}dinger equation can be autonomously be derived from
Hamilton's equations of motion; by \textquotedblleft
autonomously\textquotedblright\ I mean without recourse to any extraneous ad
hoc physical assumption. This possibility has been known for a long time by
mathematicians working in representation theory, and is an immediate
consequence of a property of the metaplectic representation of the symplectic
group. The punchline is that Schr\"{o}dinger's equation is equivalent to
Hamilton's equations of motion!

\item The uncertainty principle of quantum mechanics is already present, at
least at a formal level, in classical mechanics in its Hamiltonian
formulation. This is a consequence of a difficult result from symplectic
topology, known as Gromov's non-squeezing theorem. That theorem, which is a
considerable refinement of Liouville's theorem on conservation of phase-space
volume under canonical transformations, was only proved in 1985, and is
therefore not widely known by mathematicians --let alone physicists. Its
consequences have certainly not been fully exploited yet.
\end{itemize}

\section{Metaplectic Group and Schr\"{o}dinger Equation}

\subsection{The case of quadratic Hamiltonians}

Consider a system consisting of $N$ particles moving in physical 3-dimensional
space. We assume that the phase-space evolution of that system is governed by
Hamilton's equations%
\begin{equation}
\frac{dx_{j}}{dt}=\frac{\partial H}{\partial p_{j}}\text{ \ , \ }\frac{dp_{j}%
}{dt}=-\frac{\partial H}{\partial x_{j}} \label{ham}%
\end{equation}
where $x_{1},x_{2},x_{3}$ (resp. $p_{1},p_{2},p_{3}$) are the position (resp.
momentum) coordinates of the first particle, and so on. (A standard reference
for the Hamiltonian mechanics I will be using can be found in any of the
editions or re-editions of Goldstein \cite{Gold}). We now make the following
assumption on the Hamilton function $H$: it is a homogeneous quadratic
polynomial in the variables $x_{j},p_{k}$. Inn this case the flow determined
by the system of differential (\ref{ham}) is linear, and consists of linear
canonical transformations, that is of \textit{symplectic matrices}. Setting
$x=(x_{1},...,x_{3N})^{T}$, $p=(p_{1},...,p_{3N})^{T}$ this means that the
solution of (\ref{ham}) is given, at time $t$, by%
\[%
\begin{pmatrix}
x(t)\\
p(t)
\end{pmatrix}
=S_{t}%
\begin{pmatrix}
x(0)\\
p(0)
\end{pmatrix}
\]
where $S_{t}$ is a $6N\times6N$ real matrix such that
\[
S_{t}JS_{t}^{T}=S_{t}^{T}JS_{t}=J
\]
where $J$ is the standard symplectic matrix: $J=%
\begin{pmatrix}
0 & I\\
-I & 0
\end{pmatrix}
$ where $0$ and $I$ are, respectively, the zero and unity $3N\times3N$
matrices. In fact, if one writes the Hamiltonian $H$ in the form
\begin{equation}
H(x,p)=\frac{1}{2}z^{T}Mz\text{ \ , \ }z=%
\begin{pmatrix}
x\\
p
\end{pmatrix}
\label{hxp}%
\end{equation}
where $M$ is symmetric, then the matrices $S_{t}$ are explicitly given by the
exponential
\[
S_{t}=e^{tJM}.
\]
Now, the set of all matrices $S$ obeying the relations $SJS^{T}=S^{T}JS=J$
form a group, namely the well-known symplectic group; we denote it by
$\operatorname*{Sp}(6N)$. Thus, when $t$ varies the matrices $S_{t}$ will
describe a curve $\Sigma$ in the symplectic group, and that curve passes
passing the identity at time $t=0$. A fundamental fact is now that
$\operatorname*{Sp}(6N)$ has a double covering group. This covering is called
the metaplectic group; we denote it by $\operatorname*{Mp}(6N)$. It is unique,
as an abstract group. However --and this is crucial for the rest of the
discussion-- it can be realized, in infinitely many ways, as a group of
unitary operators acting on $L^{2}(\mathbb{R}^{3N})$ (the square-integrable
functions defined on configuration space $\mathbb{R}^{3N}$); these groups are
parametrized by a positive parameter and the copy $\operatorname*{Mp}%
^{\varepsilon}(6N)$ corresponding to the choice $\varepsilon$ for this
parameter will contain the Fourier-like transform $\widehat{F}^{\varepsilon}$
defined by%
\begin{equation}
\widehat{F}\psi(p)=\left(  \tfrac{1}{2\pi i\varepsilon}\right)  ^{3N/2}%
\int_{\mathbb{R}^{3N}}e^{\frac{i}{\varepsilon}p\cdot x}\psi(x)dx. \label{FT}%
\end{equation}
Let us now fix once for all $\varepsilon$. Using a standard property from the
theory of covering groups (the \textquotedblleft path lifting
property\textquotedblright), one proves that the curve $\Sigma$ induces
unambiguously a unique curve $\widehat{\Sigma}$ in $\operatorname*{Mp}%
^{\varepsilon}(6N)$ passing through the identity operator at time $t=0$; it is
the unique curve having this property and such that the projection of a point
$\widehat{S}_{t}$ of $\widehat{\Sigma}$ \textquotedblleft down to
$\operatorname*{Sp}(6N)$\textquotedblright\ is precisely the symplectic matrix
$S_{t}$. In view of what has been said before, the curve $\widehat{\Sigma}$
consists of unitary operators acting on $L^{2}(\mathbb{R}^{3N})$; let us let
$\widehat{S}_{t}$ act on a smooth square-integrable function $\psi_{0}$; this
defines a function of both position $x$ and time $t$:%
\begin{equation}
\psi(x,t)=\widehat{S}_{t}\psi_{0}(x). \label{wf}%
\end{equation}
The notion is intended to suggest that $\psi(x,t)$ might be some kind of
\textquotedblleft wavefunction\textquotedblright. This is indeed the case: one
proves (see below for a sketch of the proof) that $\psi(x,t)$ satisfies the
Schr\"{o}dinger-like equation%
\begin{equation}
i\varepsilon\frac{\partial}{\partial t}\psi(x,t)=H(x,-i\varepsilon\nabla
_{x})\psi(x,t) \label{sch1}%
\end{equation}
where $H(x,-i\varepsilon\nabla_{x})$ is the partial differential operator
obtained from the Hamilton function $H$ through the symmetrized
\textquotedblleft quantization rules\textquotedblright\ $x_{j}\longrightarrow
\widehat{x_{j}}$ (multiplication by $x_{j}$), $p_{j}\longrightarrow
\widehat{p_{j}}=-i\varepsilon\partial/\partial x_{j}$, and $x_{j}%
p_{k}\longrightarrow\frac{1}{2}(\widehat{x_{j}}\widehat{p_{k}}+\widehat{p_{k}%
}\widehat{x_{j}})$. Thus, if we choose the value of the arbitrary parameter
$\varepsilon$ to be $\hbar=h/2\pi$, then we obtain exactly Schr\"{o}dinger's
equation
\begin{equation}
i\hbar\frac{\partial}{\partial t}\psi(x,t)=H(x,-i\hbar\nabla_{x})\psi(x,t).
\label{sch2}%
\end{equation}

Admittedly, a mathematical equation is not a physical theory; to give the
equations above a physical sense, one needs an interpretational apparatus,
provided by physical considerations. Still, the equation is there; our
argument --which is solely based on mathematical arguments, and does not
invoke any physical assumption-- shows in the end that equation (\ref{sch1}),
and hence also Schr\"{o}dinger's equation (\ref{sch2}) are mathematically
equivalent to Hamilton's equations!

This observation is actually closely related to the fact that Ehrenfest's
equation%
\begin{equation}
m\frac{d^{2}\left\langle x\right\rangle }{dt^{2}}=-\left\langle \frac{\partial
V}{\partial x}(x)\right\rangle \label{ehr1}%
\end{equation}
discussed by Hartle becomes
\begin{equation}
m\frac{d^{2}\left\langle x\right\rangle }{dt^{2}}=-\frac{\partial V}{\partial
x}(\left\langle x\right\rangle ) \label{ehr2}%
\end{equation}
when the potential $V$ is quadratic; in fact one can prove that for all
quadratic Hamiltonians the time-evolution of the averages $\left\langle
x_{j}\right\rangle $ and $\left\langle p_{j}\right\rangle $ is governed by the
Hamilton equations (\ref{ham}) (formula (\ref{ehr2}) is an immediate
consequence of this statement when $H$ is of the type \textquotedblleft
kinetic energy plus potential $V$\textquotedblright.

\subsection{The Feynman integral}

Here are two very simple explicit examples; we work in spatial dimension $1$
for notational simplicity, but everything carries trivially through in higher
dimensions. Assume first that $H=p^{2}/2m$ , the free particle Hamiltonian.
Then
\[
S_{t}=%
\begin{pmatrix}
1 & t/m\\
0 & 1
\end{pmatrix}
\]
and, using general formulae for the metaplectic representation, one finds that
the solution $\psi(x,t)=\widehat{S}_{t}\psi_{0}(x)$ of the Schr\"{o}dinger
equation
\begin{equation}
i\hbar\frac{\partial}{\partial t}\psi(x,t)=-\frac{\hbar^{2}}{2m}\frac
{\partial^{2}}{\partial x^{2}}\psi(x,t) \label{schfree}%
\end{equation}
with initial datum $\psi_{0}$ is given, for $t\neq0$, by%
\begin{equation}
\psi(x,t)=\int_{\infty}^{\infty}K_{t}(x,y)\psi_{0}(y)dy \label{ker}%
\end{equation}
where the kernel function is given by%
\begin{equation}
K_{t}(x,y)=(e^{i\pi/4})^{\operatorname{sign}t}\sqrt{\frac{m}{2\pi\hbar|t|}%
}\exp\left[  \frac{i}{\hbar}\frac{m(x-y)^{2}}{2t}\right]  . \label{ker1}%
\end{equation}
Suppose next that $H$ is the harmonic oscillator Hamiltonian; for simplicity
we choose $m=\omega=1$ so that $H=\frac{1}{2}(p^{2}+x^{2})$; in this case the
solution of Schr\"{o}dinger's equation%
\begin{equation}
i\hbar\frac{\partial}{\partial t}\psi(x,t)=\frac{1}{2}\left(  -\hbar^{2}%
\frac{\partial^{2}}{\partial x^{2}}+x^{2}\right)  \psi(x,t) \label{scharm}%
\end{equation}
is given by (\ref{ker}), where the kernel is this time (for $t\neq n\pi$)
\begin{equation}
K_{t}(x,y)=i^{-[t/\pi]}\sqrt{\frac{1}{2\pi\hbar|\sin t|}}\exp\left[  \frac
{i}{2\hbar}\frac{(x^{2}+y^{2})\cos t-xy}{2\sin t}\right]  . \label{ker2}%
\end{equation}
These formulae are of course well known by quantum physicists; they can be
found for instance in Feynman--Hibbs \cite{FH} (but beware of misprints!)
where they are presented as cases where the \textquotedblleft Feynman path
integrals\textquotedblright\ can \textquotedblleft be done
exactly\textquotedblright. The rub is that the Feynman integral approach is as
string theory: it is not even wrong! A Feynman integral is an object which
does not (outside a few cases) make sense mathematically, and hence does not
exist\footnote{I admit that I am being a little bit unfair and grouchy at this
stage;\ Feynman integrals certainly have a good heuristic value in many
cases.}. What happens is that the quadratic nature of the Hamiltonians lead to
correct expressions for the \textquotedblleft actions\textquotedblright\ via
the Hamilton--Jacobi equation;\ this amounts to find exact generating
functions for the Hamiltonian (\textquotedblleft Hamilton's two-point
characteristic function). It is thus clear if one looks at the derivation of
formulae (\ref{ker1}) and (\ref{ker2}) in texts using the Feynman approach
(for instance Schulman \cite{Schulman}) that these methods only reconstruct
the metaplectic representation via a highly illegitimate legerdemain!

\subsection{How far further can we go?}

The reader will of course object that our considerations apply \textit{strictu
sensu} only to a very small class of physical systems, which are variations on
the theme of the free particle, or the harmonic oscillator. I totally agree
with this objection, especially since it is well-known that the quantum
behavior of systems with quadratic Hamiltonians is very close to the classical
behavior\footnote{One of the best studies of that kind of \textquotedblleft
classical vs. quantum\textquotedblright\ situation is --for my money--
Littlejohn's seminal paper \cite{Littlejohn}.}, as is exemplified by
Ehrenfest's theorem which reduces to the classical Hamilton equations in this
case. There is, moreover, a famous mathematical theorem, due Groenewold and
Hove that says that we cannot use the metaplectic representation to construct
solutions to Schr\"{o}dinger's equations for general Hamiltonians (the proof
is actually rather complicated; the mathematically minded reader can have a
look at the proof in Chapter 1 of Guillemin and Sternberg \cite{GS}). But this
\textquotedblleft no-go\textquotedblright\ result does not of course mean that
there is no way to derive the Schr\"{o}dinger equation from Hamilton's
equations of motion.

The first step towards such a program is easy, it is actually just a rather
straightforward extension of the quadratic case. Assume that $H$ is a
non-homogeneous polynomial of degree two in the position and momentum
variables. With the notation of (\ref{hxp}) we can write%
\[
H(x,p)=\frac{1}{2}z^{T}Mz+u^{T}z
\]
where $u$ is some given vector. The flow determined by the corresponding
Hamilton equations no longer consists of symplectic matrices, but rather of
affine canonical transformations (that is of linear symplectic transformations
followed (or preceded) by a phase-space translation). Such transformations
again form a group, the inhomogeneous symplectic group $\operatorname*{ISp}%
(3N)$ (it is the semi-direct product of the symplectic group and of the
translation group). It turns out that we can repeat the same procedure as in
the linear case, and show that there is, for every $\varepsilon>0$, a
one-to-one correspondence between continuous curves in $\operatorname*{ISp}%
(3N)$, and curves in a group of unitary operators, denoted by
$\operatorname*{IMp}^{\varepsilon}(3N)$ and called the \textit{inhomogeneous
metaplectic group}. $\operatorname*{IMp}^{\varepsilon}(3N)$ consists of
operators in $\operatorname*{Mp}^{\varepsilon}(3N)$ composed (on the left, or
the right) with the the Heisenberg operators
\[
\widehat{T}(x_{0},p_{0})\psi(x)=\exp\left[  \frac{i}{\varepsilon}(p_{0}\cdot
x-\frac{1}{2}p_{0}\cdot x_{0})\right]  \psi(x-x_{0})
\]
familiar from the Schr\"{o}dinger representation of the Heisenberg group when
$\varepsilon=\hbar$ (the dot $\cdot$ stands for the usual scalar product of
vectors). Everything now carries over mutatis mutandis, and we conclude that,
again, Hamilton's equations are mathematically equivalent to the
Schr\"{o}dinger-type equation associated with the non-homogeneous Hamiltonian
$H$.

Can we do anything similar in more general cases? Yes, we can. Without going
too much into details (our approach is here really sketchy) the procedure
works as follows. Let $f_{t}$ be the flow determined by the Hamilton equations
(\ref{ham}). We now use the following trick, called the \textquotedblleft
nearby orbit method\textquotedblright\ (see Littlejohn \cite{Littlejohn} for a
review of the method). It consists in replacing the Hamiltonian function $H$
by its Taylor series to the second order around a point $z_{t}=f_{t}(z_{0})$
where the initial point $z_{0}=(x_{0},p_{0})$ is arbitrary. One thus replaces
$H$ by the always time-dependent Hamiltonian%
\[
H_{z_{0}}(z,t)=H(z_{t})+\nabla_{z}H(z_{t})\cdot(z-z_{t})+\tfrac{1}{2}%
H^{\prime\prime}(z_{t})(z-z_{t})\cdot(z-z_{t})
\]
($H^{\prime\prime}(z_{t})$ is the Hessian matrix of $H$ calculated at $z_{t}%
$). Since $H_{z_{0}}(z,t)$ is a second degree polynomial in the position and
momentum variables, our discussion above applies, and the Hamilton equations
for $H_{z_{0}}$ define a flow $f_{z_{0,t}}$ consisting of affine symplectic
transformations, i.e. each $f_{z_{0,t}}$ is in the inhomogeneous group
$\operatorname*{ISp}(3N)$. Now we make the following observation: when $t$
varies $f_{z_{0,t}}(z_{0})$ is just $z_{t}$, the solution of Hamilton's
equations with initial datum $t$. Expressed in geometric terms this means that
every Hamiltonian trajectory comes from an affine flow (but this flow depends
each time on the initial point). This fact is well-known, and has been used
with profit to construct short-time solutions for Schr\"{o}dinger's equation
with initial datum a narrow wavepacket, by propagating the center of this
wavepacket along the classical curve (see again Littlejohn \cite{Littlejohn};
also de Gosson \cite{ESI}): for this purpose, it suffices to lift, as we did
before, the affine Hamiltonian flow to the inhomogeneous metaplectic group.
Using the theory of Gabor frames from time-frequency analysis it is then
possible to write such short-time solutions for arbitrary wavepackets; the
validity of these solutions breaks down after some time (called
\textquotedblleft Ehrenfest time\textquotedblright\ in the literature),
however there asymptotic validity for short times is sufficient to construct
exact solutions by a Lie-Trotter argument. So (up to mathematical difficulties
we do not discuss here) one ends up with wavepackets obeying Schr\"{o}dinger's equation.

\section{Uncertainty Principle and Symplectic Camel}

\subsection{Gromov's non-squeezing theorem}

Hamiltonian motions consist of canonical transformations, and are thus volume
preserving: this is Liouville's theorem, one of the best known results from
elementary statistical mechanics. Liouville's theorem is perhaps also one of
the most \emph{understated} because in addition of being volume-preserving
Hamiltonian motions have a surprising --I am tempted to say an
\emph{extraordinary}-- additional property as soon as the number of degrees of
freedom is superior to one. Let me shortly describe this property. Assume that
we are dealing with a Hamiltonian system consisting of a large number $N$ of
particles. If the points are sufficiently close to each other and in
sufficiently large number we may, with a good approximation, identify that
population with a \textquotedblleft cloud\textquotedblright\ of phase space
fluid. Suppose that this cloud is, at time $t=0$ spherical so we identify it
with a ball
\[
B(r):|x|^{2}+|p|^{2}\leq r^{2}%
\]
where $|x|^{2}=x_{1}^{2}+\cdot\cdot\cdot+x_{n}^{2}$ and $|p|^{2}=p_{1}%
^{2}+\cdot\cdot\cdot+p_{n}^{2}$. The orthogonal projection of that ball on any
plane of coordinates $x_{j},p_{k}$ will obviously be a circle with area $\pi
r^{2}$. From now one we assume that this plane is a plane of \emph{conjugate
coordinates}, that is $x_{1},p_{1}$ or $x_{2},p_{2}$, etc. Let us watch the
motion of this spheric phase-space cloud as time evolves. It will distort and
may take after a while a very different shape, while keeping constant volume
in view of Liouville's theorem. However --and this is the surprising result--
the projections of that deformed ball on any of the planes of conjugate
coordinates will never decrease below its original value $\pi r^{2}$ ! If we
had chosen, on contrary, a plane of non-conjugate coordinates (such as
$x_{1},p_{2}$ or $x_{1},x_{2}$, for example) then there would be no
obstruction for the projection to become arbitrarily small. This fact , which
is mathematical theorem proved in 1985 by Gromov \cite{Gromov}, is of course
strongly reminiscent of the uncertainty principle of \textit{quantum}
mechanics, of which it can be viewed as a \textit{classical} version!

This is in effect an extraordinary result, because it seems at first sight to
conflict with the usual conception of Liouville's theorem: according to
conventional wisdom, the ball $B(r)$ can be stretched in all directions, and
eventually get very thinly spread out over huge regions of phase space, so
that the projections on any plane could \textit{a priori} become arbitrary
small after some (perhaps very long) time $t$. In fact, one may very well
envisage that the larger the number $N$ of degrees of freedom, the more that
spreading will have chances to occur since there are more and more directions
in which the ball is likely to spread! This possibility has led to many
philosophical speculations about the stability of Hamiltonian systems. For
instance, in his 1989 book Roger Penrose (\cite{penrose}, p.174--184) comes to
the conclusion that phase space spreading suggests that `\textit{classical
mechanics cannot actually be true of our world}'\ (p. 183, l.--3). He however
adds that \textquotedblleft\textit{quantum effects can prevent this
spreading}\textquotedblright\ (p. 184, l. 9). Penrose's second observation
goes right to the point: while phase space spreading a priori opens the door
to classical chaos, quantum effects have a tendency to `tame'\ the behavior of
physical systems by blocking and excluding most of the classically allowed
motions. However, the phenomena we shortly described above show that there is
a similar taming in Hamiltonian mechanics preventing anarchic and chaotic
spreading of the ball in phase space which would be possible if it were
possible to stretch it inside arbitrarily thin tubes in directions orthogonal
to the conjugate planes.

Now, why do we refer to a \textit{symplectic camel} in the title of this
section? This is because one can restate Gromov's theorem in the following
way: there is no way to deform a phase space ball using canonical
transformations in such a way that we can make it pass through a hole in a
plane of conjugate coordinates $x_{j},p_{j}$ if the area of that hole is
smaller than that of the cross-section of that ball. Recalling now that in
\textit{Matthew} \textbf{19}(24) it is stated that

\begin{quote}
`...\emph{Then Jesus said to his disciples, `Amen, I say to you, it will be
hard for one who is rich to enter the kingdom of heaven. Again I say to you,
it is easier for a camel to pass through the eye of a needle than for one who
is rich to enter the kingdom of God}'.\textit{\footnote{Also see St Luke
\textbf{18}(25) and Mk \textbf{10}(25).}}'
\end{quote}

\noindent we see that in this case the Biblical camel is symplectic! For this
reason we will refer to Gromov's theorem and its variant just described as the
\textit{principle of the symplectic camel}.

Our discussion above was of a purely qualitative nature. It turns out that we
can do better, and produce quantitative statements using the principle of the
symplectic camel. For this purpose it is very useful to introduce the
topological notion of \textit{symplectic capacity}.

\subsection{Symplectic capacities}

Consider an arbitrary region $\Omega$ in phase space $\mathbb{R}^{6N}$; this
region may be large, or small, bounded, or unbounded. By definition the Gromov
capacity of $\Omega$ is the (possibly infinite) number $c_{\text{min}}%
(\Omega)$ which is calculated as follows: let again $B(r)$ be a phase space
ball with radius $r$, and assume first that there exits no canonical
transformation sending that ball inside $\Omega$. We will then say that
$c_{\text{min}}(\Omega)=0$. Assume next that there are canonical
transformations sending $B(r)$ in $\Omega$. The supremum $R$ of all the radii
$r$ for which this is possible is called the \textit{symplectic radius} of
$\Omega$ and we define the Gromov capacity of $\Omega$ by the formula
$c_{\text{min}}(\Omega)=\pi R^{2}$. Thus $c_{\text{min}}(\Omega)=\pi R^{2}$
means that one can find canonical transformations sending $B(r)$ inside
$\Omega$. for all $r<R$, but that no canonical transformation will send a ball
with radius larger $R$ inside that set. By its very definition we see that the
Gromov capacity is a symplectic invariant, that is
\begin{equation}
c_{\text{min}}(f(\Omega))=c_{\text{min}}(\Omega)\text{ if }f\text{ is
canonical;} \label{can}%
\end{equation}
it is obviously also monotone:
\begin{equation}
c_{\text{min}}(\Omega)\leq c_{\text{min}}(\Omega^{\prime})\text{ if }%
\Omega\text{ is a subset of }\Omega^{\prime} \label{monot}%
\end{equation}
and $2$-homogeneous under phase space dilations:%
\begin{equation}
c_{\text{min}}(\lambda\Omega)=\lambda^{2}c_{\text{min}}(\Omega) \label{lamb}%
\end{equation}
($\lambda$ a scalar; $\lambda\Omega$ consists of all points $(\lambda\Omega
x,\lambda\Omega p)$ such that $(x,p)$ is in $\Omega$). However, the most
striking property is the following: let us denote by $Z_{j}(R)$ the
phase-space cylinder based on the plane of conjugate variables: it consists of
all phase space points whose $j$-th position and momentum coordinate satisfy
$x_{j}^{2}+p_{j}^{2}\leq R^{2}.$ We have%
\begin{equation}
c_{\text{min}}(B(R))=\pi R^{2}=c_{\text{min}}(Z_{j}(R)). \label{cyl}%
\end{equation}
While the equality $c_{\text{min}}(B(R))=\pi R^{2}$ is immediate by definition
of $c_{\text{min}}$, the equality $c_{\text{min}}(Z_{j}(R))=\pi R^{2}$ is just
a reformulation of Gromov's non-squeezing theorem, and hence very a deep
property! In fact Gromov's theorem says that there is no way we can squeeze a
ball with radius $R^{\prime}>R$ inside that cylinder, because if we could then
the orthogonal projection of the squeezed ball would be greater than the
cross-section $\pi R^{2}$ of the cylinder, contradicting Gromov's theorem.
Thus $c_{\text{min}}(Z_{j}(R))\leq\pi R^{2}$. That we actually have equality
is immediate, since we can translate the ball $B(R)$ inside $Z_{j}(R)$ and
phase space translations are canonical transformations in their own right.

More generally one calls symplectic capacity any function associating to
subsets $\Omega$ of phase space a non-negative number $c(\Omega)$, or
$+\infty$, and for which the properties (\ref{can}), (\ref{monot}),
(\ref{lamb}), and (\ref{cyl}) are verified (see Hofer and Zehnder's book
\cite{hoze94} for very interesting examples. A caveat: the reading of this
book requires some expertise in topology and geometry; in \cite{ICP,Birk} I
have given a \textquotedblleft milder\textquotedblright\ --but also far less
complete-- treatment of this topic). There exist infinitely many symplectic
capacities\footnote{It seems to have become a kind of sport in symplectic
topology to invent new capacities!}, and the Gromov capacity is the smallest
of all:\ $c_{\text{min}}(\Omega)\leq c(\Omega)$ for all $\Omega$ and $c$. Is
there a \textquotedblleft biggest\textquotedblright\ symplectic capacity
$c_{\max}$? Yes there is, and it is actually constructed by using again
Gromov's non-squeezing theorem. It is defined as follows: suppose that no
matter how large we choose $r$ there exists no canonical transformation
sending $\Omega$ inside a cylinder $Z_{j}(r)$. We then write any $c_{\max
}(\Omega)=+\infty$. Suppose that, on the contrary, there are canonical
transformations sending $\Omega$ inside some $Z_{j}(r)$. and let $R$ be the
supremum of all such $r$. Then we set $c_{\max}(\Omega)=\pi R^{2}$. Using the
definition of a symplectic capacity it is not difficult to show that $c_{\max
}$ is a symplectic capacity, and that we have
\begin{equation}
c_{\text{min}}(\Omega)\leq c(\Omega)\leq c_{\max}(\Omega) \label{capmm}%
\end{equation}
for every symplectic capacity $c$. Now, having in mind the uncertainty
principle, a very nice fact is that all symplectic capacities agree on phase
space ellipsoids, and can be calculated as follows: assume that $\Omega$ is
the ellipsoid centered at $z_{0}$ and given by the condition
\begin{equation}
(z-z_{0})^{T}M(z-z_{0})\leq1 \label{ellipsoid}%
\end{equation}
where $M$ is some positive-definite matrix. Consider now the eigenvalues of
the product matrix $JM$; they are the same as those of the antisymmetric
matrix $M^{1/2}JM^{1/2}$ and are hence pure imaginary, that is of the type
$\pm i\lambda_{1},...,\pm i\lambda_{3N}$ where $\lambda_{j}>0$. We have%
\begin{equation}
c(\Omega)=\frac{\pi}{\lambda_{\max}} \label{capellipse}%
\end{equation}
where $\lambda_{\max}$ is the largest of all the positive numbers $\lambda
_{j}$

We will make use in next subsection of a weaker form of symplectic capacity,
the so-called linear symplectic capacity $c_{\text{lin}}$ (it should actually
be rather called an \textquotedblleft affine\textquotedblright\ capacity, but
we are complying with usage). It is defined exactly as the Gromov capacity
$c_{\text{min}}$ but one restricts oneself to the use of only affine
symplectic transformations: $c_{\text{lin}}(\Omega)=\pi R^{2}$ is thus the
supremum of all numbers $\pi r^{2}$ such that there exists an affine
transformation in $\operatorname*{ISp}(3N)$ sending the ball $B(r)$ inside
$\Omega$. The properties of $c_{\text{lin}}$ are quite similar to those of an
arbitrary symplectic capacity, except that we must replace the symplectic
invariance formula (\ref{can}) by the weaker condition%
\[
c_{\text{lin}}(f(\Omega))=c_{\text{lin}}(\Omega)\text{ if }f\text{ is in
}\operatorname*{ISp}(3N).
\]
It turns out that if $\Omega$ is an ellipsoid then $c_{\text{lin}}(\Omega)$ is
again given by formula (\ref{capellipse}); thus all symplectic capacities,
including the linear one agree on ellipsoids.

\subsection{The Heisenberg uncertainty principle}

Assume that we have a cloud of phase space points concentrated in some bounded
region $\Omega$; we do not assume that this cloud is spherical as we did when
we discussed Gromov's theorem, but just that it is a convex set. This is of
course always feasible by choosing for $\Omega$ the convex hull of the cloud,
that is, the intersection of all convex sets containing it. To make things
\textquotedblleft look quantum\textquotedblright\ we introduce Planck's
constant $h=\pi\hbar$ and assume that the linear symplectic capacity of
$\Omega$ is $c_{\text{lin}}(\Omega)\geq\frac{1}{2}h$ (this is a pedagogical
trick intended to create a surprise effect; we could have used as well
$\varepsilon$ instead of $\hbar$ as in the first part of this paper: $\hslash$
and $h$ are just positive parameters having a priori no physical meaning). We
now make the following remark: the convexity of $\Omega$ implies that there
exists a unique ellipsoid $\mathcal{J}_{\Omega}$ contained in $\Omega$ and
having maximal volume among all other ellipsoids contained in $\Omega$. It is
called the John ellipsoid; its existence was proven by Fritz John \cite{jo48}
in 1948 (see the nice paper \cite{Ball} by Ball for an extension of that
result). I claim that we have $c_{\text{lin}}(\mathcal{J}_{\Omega})\geq
\frac{1}{2}h$. Suppose in fact that this is not the case: $c_{\text{lin}%
}(\mathcal{J}_{\Omega})<\frac{1}{2}h$. Then, by definition of $c_{\text{lin}}%
$, there does not exist any canonical transformation (affine, or not) sending
the ball $B(\sqrt{\hbar})$ inside $\mathcal{J}_{\Omega}$. But then there can
be no affine symplectic transformation sending $B(\sqrt{\hbar})$ inside
$\Omega$, because of the $2$-homogeneity property, and this contradicts the
assumption $c_{\text{lin}}(\Omega)\geq\frac{1}{2}h$. \ Since $\mathcal{J}%
_{\Omega}$ is an ellipsoid, we can find a positive-definite $6N\times6N$
matrix $\Sigma$ such that $\mathcal{J}_{\Omega}$ consists of all phase space
points $z=(x,p)^{T}$ satisfying the condition
\begin{equation}
\frac{1}{2}z^{T}\Sigma^{-1}z\leq1. \label{john}%
\end{equation}
The notation suggests that $\Sigma$ can be viewed as a statistical covariance
matrix, so let us write it in the block-matrix form%
\[
\Sigma=%
\begin{pmatrix}
\Sigma_{XX} & \Sigma_{XP}\\
\Sigma_{PX} & \Sigma_{PP}%
\end{pmatrix}
\]
where the blocks $\Sigma_{XX},$ $\Sigma_{XP}=\Sigma_{PX}^{T}$, and
$\Sigma_{PP}$ are $3N\times3N$ matrices, which we find pleasant to write as
$\Sigma_{XX}=(\operatorname*{Cov}(X_{j},X_{k}))_{j,k}$, $\Sigma_{XP}%
=\Sigma_{XP}^{T}=(\operatorname*{Cov}(X_{j},P_{k}))_{j,k}$, and $\Sigma
_{PP}=(\operatorname*{Cov}(P_{j},P_{k}))_{j,k}$. It is not difficult to prove
(see de Gosson \cite{go06,Birk}) that the equation (\ref{john}) characterizing
the John ellipsoid is rigorously equivalent to the set of inequalities
\begin{equation}
(\Delta X_{j})^{2}(\Delta P_{j})^{2}\geq(\operatorname*{Cov}(X_{j},P_{j}%
))^{2}+\tfrac{1}{4}\hbar^{2} \label{unc}%
\end{equation}
where $(\Delta X_{j})^{2}=\operatorname*{Cov}(X_{j},X_{j})$ and $(\Delta
P_{j})^{2}=\operatorname*{Cov}(P_{j},P_{j})$. The observant reader will
recognize here the strong form of the Heisenberg uncertainty principle, due to
Robertson \cite{Rob} and Schr\"{o}dinger\footnote{Angelow and Batoni have
translated Schr\"{o}dinger's paper in English in \cite{ang}.} \cite{sch}; it
implies of course at once the textbook inequalities $\Delta X_{j}\Delta
P_{j}\geq\tfrac{1}{2}\hbar$ if one neglects the covariances.

The reader who has had the patience to follow me so far certainly thinks that
I have been tricking him somewhere. This is not the case; as we have shown in
\cite{go06,Birk} the inequalities (\ref{unc}) are mathematically equivalent to
the statement that $c(\mathcal{J}_{\Omega})\geq\frac{1}{2}h$ for every
symplectic capacity $c$; this statement is in turn equivalent to the matrix
condition
\begin{equation}
\Sigma+\frac{i\hbar}{2}J\text{ \ is semi-definite positive} \label{sigpos}%
\end{equation}
well-known from quantum optics (see for instance \cite{na90,SMD,SSM}; we have
reviewed the result in \cite{Birk}). The proof of the equivalence between
(\ref{john}) and (\ref{unc}) simply relies on elementary linear algebra, using
formula (\ref{capellipse}) which also applies to the linear symplectic capacity.

It turns out --and \textit{this} is the important point!-- that the
inequalities (\ref{unc}) are conserved in time under Hamiltonian evolution.
Thus, if condition (\ref{unc}) is true at some initial moment, then it will be
true in the future, and was true in the past. Let us show this in the case of
linear (or affine) flows. Returning to the convex phase space region $\Omega$
considered previously we assume again that $c_{\text{lin}}(\Omega)\geq\frac
{1}{2}h$. After time $t$ this region will be a new convex set $\Omega_{t}$
with same symplectic capacity (because $c_{\text{lin}}$ is invariant under
symplectic affine flows). We thus have $c_{\text{lin}}(\Omega_{t})\geq\frac
{1}{2}h$. It now suffices to consider the John ellipsoid $\mathcal{J}%
_{\Omega_{t}}$, and to introduce the corresponding covariance matrix
\[
\Sigma_{t}=%
\begin{pmatrix}
\Sigma_{XX,t} & \Sigma_{XP,t}\\
\Sigma_{PX},t & \Sigma_{PP,t}%
\end{pmatrix}
\]
leading to the uncertainty inequalities (\ref{unc}) at time $t$.

The generalization to arbitrary Hamiltonian flows goes along the same lines,
but is a little bit harder. The main observation is that a generic Hamiltonian
flow does not preserve the convexity and one can thus not in general associate
to $f_{t}(\Omega)$ a John ellipsoid; however this difficulty can be bypassed
by observing that the convex hull of $f_{t}(\Omega)$ indeed contains such an
ellipsoid. We will give a detailed study of the general case in a forthcoming work.

The reader should perhaps not be too surprised by the emergence of the
uncertainty principle from classical considerations. It seems today
sufficiently well-known that the uncertainty principle is really not enough
for characterizing a quantum state (except in the Gaussian case). In a recent
paper de Gosson and Luef \cite{golu1} have discussed this fact from a
mathematical point of view; our reflections were inspired by a previous paper
by Man'ko et al. \cite{manko}.

\subsection{A topological reformulation of the uncertainty principle?}

These \emph{mathematical} facts tend to show --to paraphrase what Basil Hiley
wrote in the foreword to my book \cite{ICP}-- that it is as if
\textquotedblleft... the uncertainty principle has left a "footprint" in
classical mechanics, and conversely\textquotedblright. They suggest that,
perhaps, the most general formulation of the uncertainty of quantum mechanics
could be topological. For instance one could envisage that phase space is
coarse-grained, not by cubic cells with volume $h^{3N}$ as is customary in
statistical mechanics, but rather by arbitrary cells $\mathcal{B}$ with
symplectic capacity $c(\mathcal{B)=}\frac{1}{2}h$ (for some, or maybe every,
symplectic capacity $c$). I have called such cells \textquotedblleft quantum
blobs\textquotedblright\ elsewhere; in \cite{golett1} I actually showed that
the consideration of quantum blobs as the finest possible coarse-graining can
be applied to all quantum systems with completely integrable classical
counterpart to recover the ground level energy. My attempts to use these
quantum blobs to also get the excited states have failed until now. Probably
some refinement of Gromov's non-squeezing theorem might be needed. Perhaps,
symplectic packing techniques as exposed in Schlenk's book \cite{schlenk}.
Another very appealing possibility would to use techniques from a
generalization of symplectic geometry, known as contact geometry. That this
approach might be promising is clear from the paper \cite{polt} by Elisahberg
et al. where my consideration of \textquotedblleft small
ellipsoids\textquotedblright\ is taken up from this point of view.

\section{Concluding Remarks}

With some afterthoughts the facts which we have exposed are not so surprising,
after all. One should not forget that quantum mechanics (at least in its
Schr\"{o}dinger formulation) is built from the very beginning on classical
(Hamiltonian) mechanics. The operator $H(x,-i\hbar\nabla_{x})$ appearing in
the Schr\"{o}dinger equation is not pulled out of thin air; it is obtained
using some \textquotedblleft quantization rule\textquotedblright\ from a very
classical object, namely the Hamiltonian function. Quantum mechanics appears
from this viewpoint as a refinement of Hamiltonian mechanics; to support this
claim (already made by Mackey \cite{Mackey} some years ago, although in a
different context) it suffices that the variant of quantum mechanics known as
\textit{deformation quantization}.

Of course, these facts do not mean that there is no such thing as
\textquotedblleft true\textquotedblright\ quantum mechanics! Planck's constant
plays, as a physical constant, already a primordial role in the understanding
of what a mixed state is. There is a very interesting notion, that of Wigner
spectrum, due to Narcowich (see \cite{na90} and the references therein). The
Wigner spectrum allows to characterize those self-adjoint operators
$\widehat{\rho}$ with trace one which really are mixed quantum states (a
number $\varepsilon$ is in the Wigner spectrum of $\widehat{\rho}$ if, when
one replaces $h$ by $\varepsilon$, the operator $\widehat{\rho}$ remains
semipositive-definite). It has recently been shown by Dias and Prata
\cite{dipr} that the only pure states having full Wigner spectrum $[0,h]$ are
Gaussian states. This fact, which is not a posteriori so surprising because
Gaussians are the quantum equivalent of phase space points, indicates that in
general one cannot let vary Planck's constant without risking inconsistencies.

Perhaps all this could be understood from the standpoint exposed in Bohm and
Hiley \cite{BH}; the ideas of \textquotedblleft implicate
order\textquotedblright\ exposed there are philosophically quite appealing. I
will not discuss such a perspective here, if only because of lack of competence.

\end{document}